\documentclass[letterpaper,oneside,12pt]{article}

\usepackage{graphicx}
\usepackage[body={17.5cm, 22cm},right=2.2cm]{geometry}
\usepackage{amssymb}

\newcommand\ee{\end{equation}}
\newcommand\be{\begin{equation}}
\newcommand\eea{\end{eqnarray}}
\newcommand\bea{\begin{eqnarray}}
\newcommand\mpl{M_{\rm Pl}}

\newcommand\de{\partial}
\newcommand{\sfrac}[2]{{\textstyle\frac{#1}{#2}}}

\begin{document}

\begin{flushright} 
{\footnotesize HUTP-05/A0020\\
  HD-THEP-05-09\\
  UCB-PTH-05/14\\
  LBNL-57558\\}  
\end{flushright}
\vspace{5mm}
\begin{center}
  {\LARGE \bf Ghosts in Massive Gravity}\\[1cm]
  {\large Paolo Creminelli$^{\rm a}$, Alberto Nicolis$^{\rm a}$,
    Michele Papucci$^{\rm b}$, and Enrico Trincherini$^{\rm c}$}
  \\[0.5cm]
 
  {\small
    \textit{$^{\rm a}$ Jefferson Physical Laboratory, \\
      Harvard University, Cambridge, MA 02138, USA}}
 
  \vspace{.2cm}
 
  {\small \textit{$^{\rm b}$ Department of Physics, University of California,
      Berkeley and Theoretical Physics Group, Ernest Orlando Lawrence
      Berkeley National Laboratory, Berkeley, CA 94720, USA}}

  \vspace{.2cm}

  {\small \textit{$^{\rm c}$ Institute for Theoretical Physics,\\
      Heidelberg University, D-69120 Heidelberg, Germany }}

\end{center}
\vspace{.8cm}

\hrule \vspace{0.3cm} 
{\small  \noindent \textbf{Abstract} \\[0.3cm]
\noindent
In the context of Lorentz-invariant massive gravity we show that classical solutions around heavy sources
are plagued by ghost instabilities. The ghost shows up in the effective field theory at huge
distances from the source, much bigger than the Vainshtein radius. Its
presence is independent of the choice of the non-linear terms added to the Fierz-Pauli Lagrangian.
At the Vainshtein radius the mass of the ghost is of order of the inverse radius, so that the
theory cannot be trusted inside this region, not even at the classical level.

\vspace{0.5cm}  \hrule


\section{Introduction}

In recent years there has been renewed interest in the possibility of giving a mass to the graviton. 
This idea belongs to a broader class of proposals for modifying gravity at large distances. Besides their
theoretical interest, these models could be
phenomenologically relevant as possible  alternatives to dark matter and dark energy.
In this paper we reconsider the issue of the {\it range of validity} of massive
gravity; in particular we concentrate on the stability of classical solutions around massive sources.
 
The problem we want to address has a long history. Already in the
first paper \cite{Fierz:1939ix} Fierz and Pauli observed that the mass term 
must be of the form $m_g^2(h^2-h_{\mu\nu}h^{\mu\nu})$, otherwise a ghost appears in the
spectrum, besides the 5 degrees of freedom of the massive spin-2 graviton.
A different structure would result in an instability at an energy scale $\sim m_g$.
Unfortunately Boulware and Deser showed that this additional degree of freedom 
propagates when nonlinearities in the action are taken
into account \cite{BD}. However,  from an effective field theory point of view this is not necessarily a problem,
until one specifies the scale at which the ghost shows up, {\em i.e.}~its mass. If this scale is above
the UV cutoff $\Lambda$ of the effective theory this instability can be consistently disregarded.

On the other hand, non-linearities of the classical theory are also the solution to the
problem raised by van Dam, Veltman, and Zakharov (vDVZ)
\cite{vanDam:1970vg,zakharov}: in the linearized theory predictions are
not continuous in the limit $m_g \rightarrow 0$, because the helicity-0 component of the 
graviton does not decouple from matter.
However, Vainshtein \cite{Vainshtein:1972sx} observed that the
vDVZ discontinuity might not be relevant for macroscopic sources because
the linearized approximation around a source of mass $M_*$ breaks down
at a distance $R_V = (M_* M_P^{-2} m_g^{-4})^{1/5}$ which diverges for
$m_g \rightarrow 0$. 
Classical nonlinearities become important, and the full non-linear solution could be in
perfect agreement with experiments.
This is still an open issue in massive gravity, but the Vainshtein effect has been shown to work 
in a closely related model, the DGP model \cite{DDGV,gruzinov,porrati}.

More recently massive gravity has been reconsidered in the effective
field theory language, which provides a systematic framework for dealing with
quantum effects \cite{Arkani-Hamed:2002sp}. 
For this purpose it is useful to restore the broken diffeomorphism invariance
by introducing a set of Goldstone bosons. 
With this method it is easy to see that the scalar longitudinal component of the 
graviton becomes strongly coupled at a very low energy scale, much lower than what 
naively expected by analogy with the spin-1 case. In the Fierz-Pauli theory
the strong interaction scale is $\Lambda_5 \sim (m_g^4 M_P)^{1/5}$, which for $m_g$ of 
order of the present Hubble parameter is $(10^{11}\:{\rm km})^{-1}$. 
By adding to the Fierz-Pauli Lagrangian a set of properly tuned interactions of the form
$h_{\mu\nu}^n$ the cutoff can be raised up to $\Lambda_3 \sim (m_g^2 M_P)^{1/3} 
\sim (1000\: {\rm km})^{-1}$ \cite{Arkani-Hamed:2002sp,schwartz}.

In both cases the theory seems to lose predictivity at very large
distances. 
For instance one can wonder how this strong coupling affects
the gravitational potential generated by an astrophysical source. 
Apparently the potential is uncalculable at distances smaller than $1000$ km;
but in principle this could not be the case. After all the strong
coupling takes place in the Goldstone sector, and inside the Vainshtein
radius, if the Vainshtein effect applies, one expects the Goldstone to
give a negligible correction to the Newtonian potential. Whatever
quantum effects take place at the cutoff distance, they could be
sufficiently screened from experiments. Nevertheless, without further assumptions, from an effective
theory point of view one should include in the Lagrangian all the possible operators
allowed by the symmetries and weighted by the cutoff. 
In this case the effective theory loses predictivity at a much larger length scale:
these higher dimension operators all become important at a huge distance from the source when 
evaluated on the classical solution.
In the improved theory with cutoff $\Lambda_3$ this happens at the corresponding Vainshtein radius  $R_V \sim (M_* M_P^{-2} m_g^{-2})^{1/3}$. 
For the sun $R_V$ is $\sim 10^{16}$ km. 
This means that we are unable to compute the gravitational potential at distances shorter than $R_V$ without a UV completion. 
As a consequence there is no range of distances where nonlinear effects can 
be reliably computed in the effective field theory. 
If we restrict to the original theory with cutoff $\Lambda_5$ the
situation is even worse. In this case the infinite tower of higher dimension operators become important 
at a distance which is parametrically larger than the corresponding
Vainshtein radius \cite{Arkani-Hamed:2002sp}.

The picture looks very similar to the DGP model \cite{Dvali:2000hr}, where the same problems 
have been pointed out \cite{Luty:2003vm}.
In the DGP model our world is the 4D boundary of an infinite 5D spacetime. 
Gravity is described by a standard 5D Einstein-Hilbert action with Planck mass $M_5$ in the bulk and by an additional 
4D Einstein-Hilbert action localized on the boundary, with a much larger Planck mass $M_4$.
The resulting Newton's law is 4-dimensional below the critical length scale $L_{\rm DGP} = M_4^2/M_5^3$
and 5-dimensional at larger distances.
From the 4D viewpoint
there is a scalar degree of freedom, the brane bending mode $\pi$, whose dynamics
is closely related to that of the longitudinal Goldstone boson $\phi$ of massive gravity.
In particular strong interactions show up in the $\pi$ sector at a tiny energy scale $\Lambda_{\rm DGP}
\sim (M_P/ L_{\rm DGP}^2)^{1/3}\sim (1000 \:{\rm km})^{-1}$ (taking
$L_{\rm DGP}$ of order of the present Hubble horizon $H_0^{-1}$).
Also if one includes in the Lagrangian all possible operators
allowed by the symmetries and suppressed by $\Lambda_{\rm DGP}$, around a heavy source they all  become
important at the Vainshtein radius $R_{V} \sim (M_* M_P^{-2} L_{\rm DGP}^2)^{1/3}$ when evaluated on the classical solution.

However all these difficulties depend on assumptions about the UV
completion. In the DGP model one can consistently
assume a UV completion such that the effective theory is predictive down
to distances significantly shorter than $1/\Lambda_{\rm DGP}$. For instance on the
surface of the earth the cutoff can be pushed up to $\sim {\rm cm}^{-1}$, 
not far from the smallest length scale $\sim 100 \: \mu{\rm m}$ at which gravity has been experimentally tested. 
Of course a necessary requirement for this to be possible is the consistency of the classical theory: in particular
in DGP no classical instability develops in the $\pi$ sector for all relevant astrophysical sources and for a large class 
of cosmological solutions \cite{Nicolis:2004qq}.
In this paper we want to study whether the same stability properties hold for massive gravity. 
This is a basic consistency requirement one has to satisfy
before analyzing the theory at the quantum level and
looking for a mechanism, analogous to that working in DGP, that can make the theory
predictive in a phenomenologically interesting range of scales.

The most convenient way  to study the dynamics of the theory is to use the Goldstone formalism, that we 
review in section \ref{review}; for our purposes the main interesting features of the model are 
encoded in the Lagrangian of the longitudinal component $\phi$ of the Goldstone vector. 
If we start with a Fierz-Pauli mass term, the dominant interactions for this scalar degree of freedom are cubic 
self-couplings with 6 derivatives of the form $(\de^2 \phi)^3$. 
In the presence of a macroscopic source the Goldstone gets a non-trivial configuration $\Phi(x)$; in order to study 
the stability of such a solution it is necessary to expand the action at quadratic order in the fluctuations 
around $\Phi(x)$. 
It is evident that in general, because of the cubic self-coupling, the fluctuations will get a higher-derivative 
kinetic term. As we discuss in section \ref{higher}, this signals the presence of a ghost-like instability
already at the classical level.
In DGP this does not happen: although the $\pi$ cubic self-coupling has 4 derivatives, its tensorial 
structure is such that fluctuations around a background get only a 2-derivative kinetic term
\cite{Luty:2003vm,Nicolis:2004qq}.
Unfortunately, this does not work for the Fierz-Pauli theory. 
Still, we have a large freedom in choosing the non-linear extension of FP, and one can wonder if 
it is possible to cancel all higher derivatives terms and end up with a ghost-free theory.
Sections \ref{ghosts} and \ref{unavoidable} contain the answer: despite the freedom we have, {\it ghost-like 
instabilities are unavoidable}. 

In the Goldstone language it is also easy to compute the scale at which this instability appears. 
Even in the most favorable setup in which the cutoff is $\Lambda_3$, the ghost enters in the effective 
field theory at distances from the source parametrically larger than the (already huge) Vainshtein radius $R_V$.
Furthermore, when in approaching the source we reach $r=R_V$ the mass of the ghost has dropped
to $1/R_V$! This means that in no way the theory can be extrapolated inside the Vainshtein radius.

The last part of the paper is devoted to discuss how the sickness of the theory is interpreted in the unitary gauge. 
Clearly we expect the ghost we found to be the troublesome sixth degree of freedom. 
With the Fierz-Pauli mass term, at quadratic level this degree of freedom does not appear because the trace of the 
Einstein equations gives a constraint instead of a propagating equation. 
In section \ref{unitary} we show that this equation becomes dynamical in the presence of a curved background; 
we qualitatively estimate in this simple case the mass of this new excitation and the result agrees with the 
mass we find for the ghost in the Goldstone computation.
Then, in section \ref{ADM}, we show in the Hamiltonian formalism that
there exists no non-linear extension of the Fierz-Pauli theory that can forbid the propagation of 
the sixth mode in the presence of a slightly curved background.  
The analysis in the unitary gauge is powerful for counting the number of degrees of freedom, but, 
unlike the Goldstone analysis, it says nothing about the typical scales of these modes. 
Also in order to address stability issues one should study the positivity of the Hamiltonian.
This in general is difficult, and the analysis has been carried out by Boulware and Deser for 
non-linear extensions of the form $f(h_{\mu\nu} h^{\mu\nu} - h^2)$ \cite{BD}. 
On the contrary the Goldstone analysis concentrates from the very beginning on the strong interacting
degree of freedom: all the interesting and troublesome features of the theory are encoded in 
the dynamics of a single scalar field. This enormously simplifies the analysis.

Recently it has been realized that massive gravity models with Lorentz violating
mass terms can be significantly `healthier' than the traditional Lorentz-invariant theory;
it particular they can avoid the vDVZ discontinuity and the strong coupling problem, and 
they can be free of ghosts \cite{ACLM,rubakov,dubovsky}.
In this paper we stick to the Lorentz-invariant massive gravity theory.


\section{Ghosts from higher derivative kinetic terms}\label{higher}

Let us first be very specific about why higher derivative kinetic
terms give rise to ghost-like instabilities.  Take for instance a
massless scalar field $\phi$ with Lagrangian density (note that we are using the $(-,+,+,+)$ signature!)
\be \label{4d_L}
{\cal L} = -\sfrac12(\de \phi)^2 + \frac a {2 \Lambda^2} (\Box \phi)^2
- V_{\rm int}(\phi)\; , 
\ee 
where $\Lambda$ is some energy scale,
$a=\pm 1$, and $V_{\rm int}$ is a self-interaction term.  We show
that, independently of the sign of the second term, the system is
plagued by ghosts.  To do so we want to reduce to a purely
two-derivative kinetic Lagrangian, from which we know how to read the
stability properties of the system.  We therefore introduce an
auxiliary scalar field $\chi$ and a new Lagrangian 
\be \label{mixed}
{\cal L}' = -\sfrac12(\de \phi)^2 - a \: \de_\mu \chi \de^\mu \phi -
\sfrac12 a \: \Lambda^2 \: \chi^2 - V_{\rm int}(\phi)\; , 
\ee 
which reduces exactly to ${\cal L}$ once $\chi$ is integrated out.  ${\cal
  L}'$ is diagonalized by the substitution $\phi = \phi' - a \chi$. We
get 
\be \label{chi} 
{\cal L}' = -\sfrac12 (\de \phi')^2 + \sfrac12
(\de \chi)^2 - \sfrac12 a \: \Lambda^2 \: \chi^2 - V_{\rm
  int}(\phi',\chi)\; , 
\ee 
which clearly signals the presence of a
ghost: $\chi$ has a wrong-sign kinetic term. Notice in passing that
$\chi$ can also be a tachyon, for $a=-1$: in this case $\chi$ has
exponentially growing modes. But let us neglect this possibility and
concentrate on the ghost instability, which is unavoidable. A ghost,
unlike a tachyon, is not unstable by itself: its equation of motion is
perfectly healthy at the linear level, and does not admit any
exponentially growing solution. The problem is that its Hamiltonian is
negative, so that when couplings to ordinary `healthy' matter are
taken into account (the potential term in our example above) the
system is unstable: with zero net energy one can indiscriminately
excite both sectors, and this exchange of energy happens spontaneously
already at classical level. In a quantum system with ghosts in the
physical spectrum this translates into an instability of the vacuum.
The decay rate is UV divergent due to an infinite degeneracy of the
final state phase space.  It is not clear how to cutoff this
divergence in a Lorentz invariant way \cite{riccardo}.

However the situation is not as bad as it seems: our ghost $\chi$ in
eq.~(\ref{chi}) has a (normal or tachyonic) mass $\Lambda$, so that it
will show up only at energies above $\Lambda$, {\em i.e.}~when the
four derivative kinetic term in eq.~(\ref{4d_L}) starts dominating
over the usual two derivative one.  We can consistently use our scalar
field theory eq.~(\ref{4d_L}) at energies below $\Lambda$, and
postulate that some new degree of freedom enters at $\Lambda$ and
takes care of the ghost instability.  For example, we can add a term
$-(\de \chi)^2$ to eq.~(\ref{mixed}) (for simplicity we stick to the
non-tachyonic case $a=+1$ and set $V_{\rm int} =0$), 
\be \label{L_UV}
{\cal L}_{\rm UV} = -\sfrac12(\de \phi)^2 -\de_\mu \chi \de^\mu \phi
-(\de \chi)^2 - \sfrac12\: \Lambda^2 \: \chi^2 \; .  
\ee 
This drastically changes the high-energy picture, since the resulting
Lagrangian obtained by demixing now describes two perfectly healthy
scalars, one massless and the other with mass $\Lambda$.  At the same
time, at energies below $\Lambda$ the heavy field $\chi$ can be
integrated out from ${\cal L}_{\rm UV}$, thus giving the starting
Lagrangian eq.~(\ref{4d_L}) up to terms suppressed by additional
powers of $(\de/\Lambda)^2$.  This example shows that in principle the
ghost instability can be cured by proper new physics at the scale
$\Lambda$.  In other words, eq.~(\ref{4d_L}) makes perfect sense as an
effective field theory with UV cutoff $\Lambda$.


\section{The Goldstone action}\label{review}

In this section we briefly re-derive the Lagrangian of massive gravity
along the lines of \cite{Arkani-Hamed:2002sp}, {\em i.e.}~keeping explicit
the Goldstone bosons of broken diffeomorphism invariance.

To write down a mass term for gravity, in addition to the full
dynamical metric $g_{\mu\nu}$, we have to take a reference fixed
metric: for our purposes we will take the Minkowski metric
$\eta_{\mu\nu}$. A mass term breaks invariance under
general coordinate transformations.  However, as it has been shown in
Ref.~\cite{Arkani-Hamed:2002sp}, one can always restore local
coordinate invariance by the St\"uckelberg trick: in analogy with 
massive gauge theories one introduces a set of Goldstone fields and
requires that they transform non-linearly under a local coordinate
transformation.  The fundamental object to be used for this purpose is
a symmetric tensor $H_{\alpha\beta}$, built in terms of the reference
metric, the field describing metric fluctuations
$h_{\mu\nu}=g_{\mu\nu}-\eta_{\mu\nu}$ and the four Goldstone fields
$\pi_{\mu}$,
\begin{equation}
  \label{eq:Hdef}
  H_{\alpha\beta}=h_{\alpha\beta}+\partial_{\alpha}
  \pi_{\beta}+\partial_{\beta} \pi_{\alpha}+
  \partial_{\alpha}\pi^{\gamma} \partial_{\beta} \pi_{\gamma}\;.
\end{equation}
$H_{\mu\nu}$ transforms as a covariant tensor
under local diffeomorphisms $x^{\alpha} \rightarrow
x^{\alpha}+\xi^{\alpha}$ provided that $\pi_{\alpha}$ shifts, $\pi_{\alpha}
\rightarrow \pi_{\alpha} - \xi_{\alpha}$.  As in non-abelian massive gauge theories,
since now local coordinate invariance is non linearly realized on the
$\pi$ field, a Lagrangian built using $H$ will be valid as an
effective theory and its breakdown will appear as the Goldstone sector
becoming strongly coupled at some scale $\Lambda$.  This indeed has
been shown in Ref.~\cite{Arkani-Hamed:2002sp}. It is useful to further
split the 4 $\pi^{\alpha}$ fields into a vector and a scalar as
\begin{equation}
  \label{eq:2}
  \pi_{\mu}=A_{\mu}+\partial_{\mu} \phi
\end{equation}
together with an additional hidden $U(1)$ gauge invariance for
$A_{\mu}$ under which $\phi$ shifts. Note that since $\phi$ is a
Goldstone boson under this $U(1)$ gauge symmetry and $\pi_{\mu}$ is a
Goldstone boson of broken diffeomorphism invariance $\phi$ will appear
with two derivatives in the Lagrangian, as evident from
eq.~(\ref{eq:Hdef}).  A mass term for $h_{\mu\nu}$ can be written down
in terms of $H_{\mu\nu}$ as
\begin{equation}
  \label{eq:PFterm}
  \sqrt{-g}g^{\mu\nu}g^{\alpha\beta}(a H_{\mu\alpha}H_{\nu\beta}+b H_{\mu\nu}H_{\alpha\beta})\;.
\end{equation}

Expanding $H$ using (\ref{eq:Hdef}) one easily realizes that for a
generic choice of $a,b$ there is a quadratic term in $\phi$ containing
4 derivatives. This term signals the presence of a ghost as we have
seen in Section \ref{higher}. Only for the Fierz-Pauli choice
$a=-b\equiv m_g^2M_P^2$ this four derivative term exactly cancels. In
this case $\phi$ does not have a kinetic term on its own, but only a
kinetic mixing with $h_{\mu\nu}$: $m_g^2 M_P^2 (\de_{\mu}\de_{\nu}\phi
\,h^{\mu\nu} - \Box\phi \,h)$. A conformal rescaling of the metric $h_{\mu\nu}=\hat
h_{\mu\nu}+m_g^2 \eta_{\mu\nu}\phi$ diagonalizes this mixing and
generates a small ({\em i.e.}~proportional to $m_g^2$) kinetic term for
$\phi$ besides interactions of the form $\phi (\partial^2 \phi)^n$.
The smallness of the kinetic term is the origin of the low strong
coupling scale as it enhances the $\phi$ interactions once the fields
are canonically normalized.

One can easily see that the most relevant interactions are of the form
\begin{equation}
  \label{eq:5}
  m^2 M_P^2(\partial^2 \phi)^3=\frac{(\partial^2 \phi^c)^3}{M_P m_g^{4}}
\end{equation}
where $\phi^c$ is the canonically normalized  field. These interactions saturate perturbation theory at the tiny energy scale $E\sim
\Lambda_5\equiv (m_g^4 M_P)^{1/5}$.

One can slightly improve the situation canceling these cubic
interactions by adding $H^3$ terms. Now the most relevant interactions
will be of the form $(\de^2 \phi)^4$ and this procedure can be repeated
at any order. The dominant interactions will be $(\de^2 \phi )^n$ and
once all these are canceled the theory has the cutoff $\Lambda_3
\equiv (m^2_g M_P)^{1/3}$.
In fact after this procedure the most relevant interactions are
\be 
m_g^2 M_P^2 \, (\de A)^2 (\de^2 \phi)^n 
\quad {\rm and} \quad 
m_g^2 M_P^2 \, (\hat h_{\mu\nu} + m_g^2 \eta_{\mu\nu} \phi)(\de^2 \phi)^n \;, 
\ee 
which are weighted by $\Lambda_3$ when expressed in terms of canonically normalized fields.

The interaction between matter and gravity is as usual described by the term $\sfrac12 h_{\alpha\beta} T^{\alpha\beta}$.
The Weyl transformation that demixes $h_{\mu\nu}$ from $\phi$ thus generates a
direct coupling of $\phi$ to the trace $T$ of the stress-energy tensor. 
This implies that we will have a non-trivial $\phi$ background
around any astrophysical source.  For a classical solution we will
have two relevant scales: the first is the Schwarzschild radius $R_S$,
the distance from the source at which linearized gravity breaks down,
the second is the Vainshtein radius $R_V$~\cite{Vainshtein:1972sx}
where nonlinearities for the scalar field $\phi$ become important.
For a source of mass $M_*$ this distance is equal to $1/\Lambda_5
(M_*/M_P)^{1/5}$ (which is much larger than $R_S$). At this scale the
term $(\partial^2\phi)^3$ becomes as relevant as the kinetic term
whereas all the other nonlinear terms are important only when we reach
the Schwarzschild radius so that they can be safely neglected.
Therefore the action of $\phi$ in the presence of sources is given by
\begin{equation}
  \label{eq:lagrphi}
  S=\int d^4 x \left\{ 3 \phi^c \Box
    \phi^c+\frac{1}{\Lambda_5^5}\left[(\Box\phi^c)^3-(\Box \phi^c)(\partial_{\mu}\partial_{\nu}\phi^c)^2\right]
    +\frac{1}{2M_P}\phi^c T\right\} \; ;
\end{equation}
the structure of the trilinear terms can be changed by adding non-linear interactions to
the Fierz-Pauli Lagrangian.

The situation remains qualitatively unchanged even if the first $N$
$(\de^2 \phi)^n$ interactions are tuned to zero. The first
non-vanishing term, $(\de^2 \phi)^{N+1}$, will set the Vainshtein
radius, while higher order terms will become relevant again at the
Schwarzschild radius.


\section{Massive gravity in the presence of a source}\label{ghosts}

Let us consider the Lagrangian eq.~(\ref{eq:lagrphi}) in the presence
of a macroscopic source, like the Sun.  This induces a classical
background $\Phi (x)$, solution of the $\phi$ equation of motion.  To
study the stability of this solution we expand the Lagrangian at the
quadratic order in the fluctuation $\varphi \equiv \phi - \Phi$.  The
result is schematically of the form 
\be \label{L_ghost} 
{\cal
  L}_\varphi = -(\de \varphi^c)^2 + \frac {(\de^2 \Phi^c)} {{\Lambda_5}^5}
(\de^2 \varphi^c)^2 \; , 
\ee 
{\em i.e.}~the background gives a
four-derivative contribution to the $\varphi$ kinetic term.  As
discussed in sect.~\ref{higher}, this results in the appearance of a
ghost with an $x$-dependent mass 
\be \label{m_ghost} 
m^2_{\rm ghost}(x) \sim \frac{{\Lambda_5}^5}{\de^2 \Phi^c(x)} \; .  
\ee 
Remember that
we are dealing with an effective theory with a tiny UV cutoff
$\Lambda_5$, therefore we should not worry until the mass of the ghost
drops below $\Lambda_5$.  In approaching the source from far away,
this happens at a distance $R_{\rm ghost}$ from the source such that
$\de^2 \Phi^c \sim {\Lambda_5}^3$.  Unfortunately this is a huge
distance, parametrically larger than the (already huge) Vainshtein
radius $R_V$.  
In fact for a source of mass $M_*$ at distances $r \gg R_V$ the background field goes as $\Phi^c
(r) \sim (M_*/M_P) \cdot 1/r$, so that 
\be  \label{R_ghost}
R_{\rm ghost} \sim
\frac1{\Lambda_5} \left(\frac{M_*}{M_P}\right)^{1/3} \; \gg \;
R_V^{(5)} \sim \frac1{\Lambda_5} \left(\frac{M_*}{M_P}\right)^{1/5}
\;.  
\ee 
Therefore the ghost is going to show up in an extremely weak
background field, when the latter is still in its linear regime.

Inside $R_{\rm ghost}$, in the spirit of sect.~\ref{higher}, one is
forced to postulate that additional physics lighter than the local
ghost mass cures the instability, that is the cutoff must be lowered
from $\Lambda_5$ to $m_{\rm ghost}(x)$.  
A byproduct of this in general would be that interactions strengthen, 
being weighted by the new cutoff scale rather than by $\Lambda_5$.
But let us optimistically assume that, instead, the only effect of this new 
physics is to cure the ghost instability.
However, when the local ghost mass is of order of the inverse distance from the source there is no
way of proceeding further without specifying the UV completion of the
theory, since the background itself has a typical length scale of
order of the UV cutoff.  One can easily check that this happens at the
Vainshtein radius $R_V$.  {\em There is no sense in which one can
  trust the classical solution below $R_V$}. Since one can hope to
recover General Relativity only in the region inside $R_V$, where non-linear
effects can hide the scalar (Vainshtein effect), this also means that {\em
  General Relativity is nowhere a good approximation}.

Notice that in the DGP model the dominant interaction of the Goldstone
has the form $\Box \pi (\de \pi)^2$, which suggests the same problem
we are facing, as there are two derivatives acting on one of the
$\pi$'s.  Nevertheless, in the equation of motion terms with more than
two derivatives acting on a single field cancel out and one is left
with a (non-linear) second order differential equation
\cite{Nicolis:2004qq}\footnote{Equivalently, working at the level of
  the Lagrangian, one can expand the interaction term $\Box \pi (\de
  \pi)^2$ to second order in the fluctuation $\varphi$ around a
  background $\pi_b$.  The worrisome term is $\Box \varphi \de_\mu
  \varphi \: \de^\mu \pi_b$, since it has 3 derivatives acting on the
  $\varphi$'s. But by integration by parts one can shift one
  derivative from the fluctuations to the background, thus obtaining
  an ordinary 2-derivative kinetic term for the fluctuations (whose
  positivity must however be checked)
  \cite{Luty:2003vm,Nicolis:2004qq}.}.  The same cannot happen in our
case since there are too many derivatives: the contribution of the
trilinear term to the equation of motion is a sum of terms with 2
$\phi$'s and 6 derivatives. In any term there is at least one $\phi$
carrying more than two derivatives.

One is thus led to consider the possibility of eliminating the
unwanted trilinear interaction of the Goldstone by adding appropriate
cubic terms in $H_{\mu\nu}$ to the Fierz-Pauli Lagrangian
eq.~(\ref{eq:PFterm}). The three independent contractions are $H^3$,
$H (H_{\mu\nu})^2$, and $(H_{\mu\nu})^3$, where the last stands for
the cyclic contraction of the indices. These contain interaction terms
for the Goldstone of the form $(\de^2 \phi)^3$ which, for the proper
choice of coefficients, cancel the trilinear interaction of
eq.~(\ref{eq:lagrphi}).  However, in this way one introduces further
quartic interactions $(\de^2 \phi)^4$ on top of those already present
in the Fierz-Pauli mass term, because of the non-linear relation
between $H_{\mu\nu}$ and $\phi$ of eq.~(\ref{eq:Hdef}).  These are
problematic for exactly the same reason as before, and the same
problem shows up at any order: an interaction term of the form $(\de^2
\phi)^n$ evaluated around a background gives a contribution to the
equation of motion for the fluctuations with too many derivatives.
This signals the presence of a ghost instability\footnote{The reader
  could wonder if there exists a choice of coefficients such that
  terms with 4 derivatives on a single field cancel in the equation of
  motion. In this case one would be left only with 3-derivative terms
  and our conclusions should be modified. But this is not the case:
  setting to zero all 4 derivative terms leads also to the
  cancellation of those with 3 derivatives.  To see this, consider the
  most general interaction Lagrangian of $n$-th order,
  ${\cal{L}}^{(n)} = \Gamma^{\alpha_1 \beta_1 \cdots \alpha_n \beta_n}
  \: \de_{\alpha_1} \de_{\beta_1} \phi \cdots \de_{\alpha_n}
  \de_{\beta_n} \phi$, where $\Gamma$ is a tensor constructed with the
  metric $\eta_{\mu\nu}$.  Given the structure of contractions,
  without loss of generality we can choose $\Gamma^{\alpha_1 \beta_1
    \cdots \alpha_n \beta_n}$ to be symmetric under $\alpha_i
  \leftrightarrow \beta_i$ and $(\alpha_i, \beta_i) \leftrightarrow
  (\alpha_j, \beta_j)$.  Then, for symmetry reasons, the contribution
  of ${\cal{L}}^{(n)}$ to the $\phi$ equation of motion is \be
  \Gamma^{\alpha_1 \beta_1 \cdots \alpha_n \beta_n} \left[ A_n \:
    (\de_{\alpha_1} \de_{\beta_1}\de_{\alpha_2} \de_{\beta_2} \phi)
    (\de_{\alpha_3} \de_{\beta_3} \phi) + B_n \:
    (\de_{\beta_1}\de_{\alpha_2} \de_{\beta_2} \phi)
    (\de_{\alpha_1}\de_{\alpha_3} \de_{\beta_3} \phi) \right]
  \de_{\alpha_4} \de_{\beta_4} \phi \cdots \de_{\alpha_n}
  \de_{\beta_n} \phi \; , \ee where $A_n, B_n$ are combinatoric
  factors.  The four-derivative term (the first in brackets)
  identically vanishes only if the totally symmetric part of
  $\Gamma^{\alpha_1 \beta_1 \cdots \alpha_n \beta_n}$ in the first
  four indices does, $\Gamma^{(\alpha_1 \beta_1 \alpha_2 \beta_2)
    \cdots \alpha_n \beta_n}=0$.  In this case, given the symmetries
  of $\Gamma$, it is straightforward to check that $\Gamma^{\alpha_1
    (\beta_1 \alpha_2 \beta_2) \cdots \alpha_n \beta_n} =
  \Gamma^{(\alpha_1 \beta_1 \alpha_2 \beta_2) \cdots \alpha_n \beta_n}
  = 0$.  This eliminates the 3 derivative term (the second in
  brackets) as well, {\em i.e.}~eliminates the contribution of
  ${\cal{L}}^{(n)}$ to the $\phi$ equation of motion altogether.}.
Again one can check that the cutoff ({\em i.e.}~the ghost mass)
becomes of order of the inverse radius at the new Vainshtein scale
(always defined as the distance from the source at which
non-linearities become relevant). Hence one never recovers General
Relativity.

The only possibility is therefore to concentrate on theories in which
all the interactions of the form $(\de^2 \phi)^n$ are set to zero by
properly choosing infinitely many coefficients. We can look at this
procedure as an extension at non-linear order of the Fierz-Pauli
choice for the mass term which, as discussed in sect.~\ref{review},
leads to the cancellation of the $(\de^2 \phi)^2$ terms. In the next
section we prove that also in this case we cannot avoid higher
derivative kinetic terms for the fluctuations of the field $\phi$
around a non trivial background: {\em ghost-like instabilities are
  unavoidable}.

\section{Ghost instabilities are unavoidable}\label{unavoidable}

The cancellation of the $(\de^2 \phi)^n$ interactions has also been
considered as a way to raise the strong interaction scale to
$\Lambda_3= (m_g^2 M_P)^{1/3} \gg \Lambda_5$
\cite{Arkani-Hamed:2002sp,schwartz}.  In fact, after the cancellation, the
leading interactions are of the form 
\be 
m_g^2 M_P^2 \, (\de A)^2
(\de^2 \phi)^n \quad {\rm and} \quad m_g^2 M_P^2 \, (\hat h_{\mu\nu}
+m_g^2 \eta_{\mu\nu} \phi)(\de^2 \phi)^n \;; 
\ee 
when the fields are
canonically normalized all these terms are suppressed by the scale
$\Lambda_3$, while additional interactions are weighted by higher
scales.  Correspondingly the Vainshtein radius now
shrinks to $R_V^{(3)} = 1/\Lambda_3 (M_*/M_P)^{1/3}\ll R_V^{(5)}$ as
the original leading interactions have been canceled.  At this radius
all non-linear terms of the form $(\hat h_{\mu\nu} + m_g^2
\eta_{\mu\nu} \phi)(\de^2 \phi)^n$ become relevant; on the other hand
there are no terms linear in $A$ in the Lagrangian, so that $A$ is sourced neither by matter
nor by the other fields.
Interactions involving this field are therefore irrelevant for our purposes, and
can be consistently neglected.

The theory we are describing is not unique: there are different
possible choices of coefficients that cancel all the interactions
$(\partial^2 \phi)^{n}$. We can easily see why.  Let us start with the
canonical Fierz-Pauli mass term\footnote{
We use the notation $[H] = g^{\mu\nu} H_{\mu \nu}$, 
$[H^2] =  g^{\mu\nu} g^{\alpha \beta} H_{\mu \alpha} H_{\nu \beta}$, 
and its straightforward generalization to higher orders.} 
${\cal L}_2= \sqrt{-g}\,\left( [H^2]-[H]^2 \right)$; 
since $H_{\mu\nu} = h_{\mu\nu} + 2 \, \partial_{\mu} \partial_{\nu} \phi 
+ \partial_{\mu} \partial_{\alpha} \phi \, \partial_{\nu} \partial^{\alpha} \phi$ (setting
$A_\mu=0$), ${\cal L}_2$ contains $(\partial^2 \phi)^3$ interactions.
We can cancel them adding an appropriate combination of terms cubic in
$H$, ${\cal L}_3 = \sqrt{-g}\,\left( \frac{1}{2} [H][H^2]- \frac{1}{2} [H^3] \right)$. 
But at this point we can still add, with an arbitrary overall coefficient
$\alpha_3$, the expression 
\be\label{t3} 
{\cal L}_3^{\rm TD} =\sqrt{-g}\,\left( 3 [H][H^2]-[H]^3-2[H^3]\right) \; , 
\ee 
because it gives $(\partial^2
\phi)^{3}$ terms in the combination 
\be\label{cubica} 
(\Box \phi)^3 - 3 \: \Box \phi \: (\de_\mu \de_\nu \phi)^2 + 2 \: (\de_\mu \de_\nu
\phi)^3 \;, 
\ee 
which is a total derivative (hence the superscript
`TD') and thus does not contribute to the equation of motion.  Now
${\cal L}_2 + {\cal L}_3 + \alpha_3 {\cal L}_3^{\rm TD}$ contains $(\de^2
\phi)^4$ interactions and we can repeat the procedure.  Fourth order
terms are canceled by 
\be\label{lquartica} 
{\cal L}_4= \sqrt{-g}\,\frac{1}{16}
\Big[ (5+24 \alpha_3) [H^4] -(1+12\alpha_3)[H^2]^2 -(4+24\alpha_3) [H][H^3] + 12\alpha_3
[H^2][H]^2 \Big].  
\ee 
Again we have the possibility to introduce a
second arbitrary coefficient, $\alpha_4$, in front of the `total
derivative' term 
\be\label{t4} 
{\cal L}_4^{\rm TD} = \sqrt{-g}\,\left(  [H]^4
-6[H^2][H]^2+8[H^3][H]+3[H^2]^2-6[H^4]\right) \;, 
\ee 
and we can go on at
higher orders until all self-couplings are removed.  
Notice that these total derivative terms ${\cal L}_n^{\rm TD}$ are the 
higher order analogue of the Fierz-Pauli mass term: they are combinations of 
$h_{\mu\nu}$ interactions that reduce to a total derivative when expressed in terms of 
$\de_\mu \de_\nu \phi$ and therefore do not contribute to high-derivative terms in $\phi$.
One can check that there is one of such combinations per order, but for our purposes we will
need them only up to fourth order\footnote{
It is easy to check that at $n$-th order a total derivative term is given by
\be
\sum_\pi (-1)^\pi \eta^{\alpha_1 \pi(\beta_1)} \cdots \eta^{\alpha_n \pi(\beta_n)} 
\: \de_{\alpha_1} \de_{\beta_1} \phi \cdots \de_{\alpha_n} \de_{\beta_n} \phi \; ,
\ee
where the sum runs over all permutations $\pi$ of the $\beta$ indices, and $(-1)^\pi$ is the parity
of the permutation.
To prove that this combination is the only total derivative term at a given order 
assume that there are two of them. One then could construct a total derivative term that does not
contain, say, $(\Box \phi)^n$. Imposing that the contribution to the field equations is zero, it is
straightforward to show that also all the other terms vanish.
}.

We now want to see if it is possible to get a kinetic structure without ghosts for
the fluctuations around a background. Note that this theory has potentially 
dangerous terms of the form $h_{\mu\nu} (\de^2 \phi)^n$.
We could hope that the large freedom we have in the choice of
higher-order terms, parameterized by the coefficients $\alpha_3, \alpha_4,\dots$,
helps us to obtain a ghost-free theory.  
Unfortunately, this is not the case and the
main reason is that the number of possible contractions of
$H_{\mu\nu}$ grows very fast and soon we cannot cancel all the dangerous kinetic terms. 
Let us see how this works explicitly. 
We call $h_{\mu\nu}^b$ and $\phi^b$ the background fields\footnote{
Remember that the field $h_{\mu\nu}$ is the graviton {\it before} the Weyl rescaling
that demixes it from $\phi$: $h_{\mu\nu}= \hat{h}_{\mu\nu} + m_g^2 \eta_{\mu\nu} \phi$.
}, 
and we study the quadratic Lagrangian for the
scalar fluctuation $\varphi$. Actually, instead of working at the level of the Lagrangian, 
the most direct approach is to look at the equations of motion linear in the fluctuations, 
because in this case there is no ambiguity coming from integration by parts.  
We have to check whether all the terms with more than 2 derivatives on
$\varphi$ can cancel in the equations of motion.

We start with the Lagrangian terms {\it cubic} in the fields: they can
come only from ${\cal L}_2 + {\cal L}_3 + \alpha_3 {\cal L}_3^{\rm TD}$.
Only terms schematically of the form $h (\de^2 \phi)^2$ can give a
ghost; terms with a higher number of $h_{\mu\nu}$ have no more than 2
derivatives.  Using the explicit expression of the Lagrangian, it is
immediate to verify that {\em i)} the terms in the e.o.m.~with more
than 2 derivatives on $\varphi$ cancel already with $\alpha_3=0$, and {\em
  ii)} they cancel also when originating from ${\cal L}_3^{\rm TD}$
alone.  We conclude that at the cubic level in $H_{\mu\nu}$ there are
no higher derivative kinetic terms for $\varphi$ and the parameter
$\alpha_3$ can still be varied arbitrarily.

What happens with the {\it quartic} Lagrangian? Let us consider the
interactions $h_{\mu\nu} (\de^2 \phi)^3$; now we must include also
${\cal L}_4$ (eq.~(\ref{lquartica})), and for simplicity we start with
$\alpha_4=0$.  Again it is straightforward to write down the equations of
motion linear in $\varphi$.  The terms with more than two derivatives
on the fluctuation can be divided into two classes: those containing
$(h^b)^\mu {}_\mu$ and those in which $(h^b)_{\mu\nu}$ is contracted
with derivatives of $\phi$.  Either class must cancel independently of
the other, since, given the different tensor structure, there is no
possibility of cross-cancellation between the two.  In the e.o.m.~the
terms belonging to the first class automatically cancel.  They come
from a piece of the Lagrangian that is precisely $4 \alpha_3 \, [h]$ times
the `total derivative' combination eq.~(\ref{cubica}).  The
remaining dangerous kinetic terms, which belong to the second class,
come from the Lagrangian\footnote{Inside brackets with $\phi$ we indicate the matrix 
$\partial_\mu\partial_\nu\phi$. For example the term $[h \,\phi][\phi]^2$ 
should be read as $\eta^{\mu\alpha}\eta^{\nu\beta}\, h_{\mu\nu}\, \partial_\alpha\partial_\beta
\phi \, (\Box\phi)^2$.} 
\be\label{lfinale} 
(8+72\alpha_3)\: ([h\, \phi^3]
- [h \,\phi^2][\phi]) - 36 \alpha_3 \: ([h \, \phi][\phi^2] - [h \,
\phi][\phi]^2) \; .  
\ee 
Can we add now $\alpha_4 {\cal L}_4^{\rm TD}$
(eq.~(\ref{t4}))? Yes, since this does not reintroduce terms of the
first class ({\em i.e.}, containing $(h)^\mu{}_\mu$) in the equations
of motion.  Then, can we choose $\alpha_4$ to get rid of the contributions
coming from eq.~(\ref{lfinale})?  Unfortunately the answer is
negative. In fact, expanding $\alpha_4 {\cal L}_4^{\rm TD}$, 
\be 
\alpha_4 {\cal
  L}_4^{\rm TD} \supset -192\alpha_4 \: ([h \, \phi^3] - [h \,
\phi^2][\phi]) + 96 \alpha_4 \: ([h \,\phi][\phi^2]-[h \, \phi][\phi]^2)
\;, 
\ee 
one immediately sees that either the first or the second pair
of terms in eq.~(\ref{lfinale}) (but not both) can be canceled by
properly choosing $\alpha_3$ and $\alpha_4$.  The other pair gives a non-zero,
four derivative contribution to the equation of motion for the
fluctuation $\varphi$.  The number of possible tensor structures is
bigger than the freedom we have in the Lagrangian.  This completes the
proof that {\it massive gravity around a generic background cannot
  have a purely two-derivative kinetic term for $\varphi$}.

In the $\Lambda_3$ theory the local mass of the ghost goes as
\be
m^2_{\rm ghost}(x) \sim \frac{{\Lambda_3}^6}{\Phi^c(x) \, \de^2 \Phi^c(x)} \; ,
\ee
so that the ghost enters in the effective theory at a distance from the source much bigger 
than the Vainshtein radius,
\be
R_{\rm ghost} \sim
\frac1{\Lambda_3} \left(\frac{M_*}{M_P}\right)^{1/2} \; \gg \;
R_V^{(3)} \sim \frac1{\Lambda_3} \left(\frac{M_*}{M_P}\right)^{1/3} \;.
\ee
These results should be compared with their analogues in the $\Lambda_5$ theory, eqs.~(\ref{m_ghost}) and
(\ref{R_ghost}).
Again at the Vainshtein scale the mass of the ghost is of order of the inverse
radius and we cannot proceed further towards the source.

There is a final subtlety that needs to be addressed. Is the presence
of a 4-derivative kinetic term enough to claim that there is a ghost?
After all, the argument of sect.~\ref{higher} strictly applies only to
a Lorentz-invariant background. It is clear that if the derivatives
acting on the fluctuation $\varphi$ are contracted with a background
tensor field different from $\eta_{\mu\nu}$ the situation can be very
different. For instance, if the quadratic Lagrangian for $\varphi$
involves 4 space derivatives but only 2 time derivatives there is no
room for an independent propagating extra scalar ($\chi$, in the
language of sect.~\ref{higher}), and thus there is no ghost, provided
that the $\dot \varphi^2$ term has the healthy sign.  Indeed, in most
astrophysical situations macroscopic sources have non-relativistic
velocities $v \ll 1$; the background $\phi^b$ field they generate is
therefore essentially constant in time, its time derivatives being
suppressed by positive powers of $v$ with respect to its spatial
gradients.  This, in a term like $\de^\mu \de^\nu \phi^b \, \de_\mu
\de_\rho \varphi \de^\rho \de_\nu \varphi$ for instance, can suppress
the magnitude of terms with 4 time derivatives. Although this is a
parametric increase of the ghost mass, for typical astrophysical
velocities $v \sim 10^{-4} - 10^{-3}$ the regime of validity of the
theory is not significantly widened: inside the Vainshtein radius the
theory breaks down at $r \sim R_V^{(5)} v^{4/5} \sim (10^{-3} - 10^{-2})
R_V^{(5)}$, where the ghost mass becomes of order of the inverse
radius.


\section{Unitary gauge description}\label{unitary}

In the previous section we argued in the Goldstone language
that Fierz-Pauli massive gravity (together with all its infinitely
many higher order extensions) is unavoidably plagued by ghosts around
a tinily curved background.
We now want to see how the sickness of the theory is interpreted in
the unitary gauge.
In order to take into account the presence of a (slightly) curved
background we need to consider the field equations at second order in
$h_{\mu\nu}$,
\be  \label{einstein}
G_{\mu\nu} + m^2_g [a \, h_{\mu\nu} + b \, h \eta_{\mu\nu} +
{\cal O}(h_{\mu\nu} ^2)] = 0 \; , 
\ee 
where the quadratic terms come both from the mass term (that we take generic for the moment)
and from additional higher order interactions.
As it is well known, the invariance of the Einstein-Hilbert action under
diffeomorphisms $g_{\mu\nu} \to g_{\mu\nu} +
\nabla_\mu \epsilon_\nu + \nabla_\nu \epsilon_\mu$, 
\be 
\delta S_{\rm EH} = 2 \int \! d^4 x \:\frac{\delta S_{\rm EH}}{\delta
  g_{\mu\nu}} \nabla_\mu \epsilon_\nu = 
- 2 \int \! d^4 x \sqrt{-g}
\:\epsilon_\nu \: \nabla_\mu \left( \frac{1}{\sqrt{-g}}\frac{\delta
    S_{\rm EH}}{\delta g_{\mu\nu}} \right) = 0 \; ,  
\ee 
implies the contracted Bianchi identities $\nabla^\mu G_{\mu\nu}=0$.
As a consequence, from eq.~(\ref{einstein}) we get the four constraints
\be \label{4_constraints_curved} 
\nabla^\mu [a \, h_{\mu\nu} + b \, h \eta_{\mu\nu} + {\cal O}(h_{\mu\nu} ^2)] = 0 \; ,  
\ee 
which reduce to six the number of propagating components of $h_{\mu\nu}$.
The presence of these constraints is ensured by the gauge-invariance of the `kinetic' action.
This is in complete analogy to what happens in the theory of a
massive vector particle, where the gauge invariance of the kinetic
term implies the constraint $\de_\mu A^\mu = 0$, thus reducing to
three the number of propagating degrees of freedom.

If we now restrict to a linear analysis,
for the particular Fierz-Pauli choice of the mass term ($b = -a$)
we have an additional constraint equation. In fact the linearized Einstein tensor satisfies
\be \label{trace_G}
G^{\ell \, \mu} {}_\mu = \de^\mu \de^\nu (h_{\mu\nu} - h
\eta_{\mu\nu}) \; , 
\ee 
so that eq.~(\ref{4_constraints_curved}) forces $G^\mu {}_\mu$ to vanish at linear order, 
and thus the trace of the Einstein equations eq.~(\ref{einstein}) becomes a constraint for the trace mode, 
\be \label{constraint_trace} 
h = 0 \; .  
\ee 
In the end one is left with
five propagating degrees of freedom, the correct number for a massive
spin-2 particle. However, it is clear that this last constraint is
fundamentally different from the previous four, since, unlike them, is
not ensured by any symmetry, but it is based on a precise tuning in
the structure of the mass term and on the identity
eq.~(\ref{trace_G}), valid at linear order. It is then natural to
expect that it does not survive in a curved background.  The ghost we
found in the Goldstone language is nothing but this hidden sixth mode
that, although constrained in flat space, starts propagating around a
non-zero background.  Let us check that the energy scale at which the
additional mode appears is indeed the same in the two descriptions.

At quadratic order $G^\mu {}_\mu$ will not vanish anymore on the equations of
motion, instead it will be of the form $G^\mu {}_\mu \sim \de^2 \,
h_{\mu\nu}^2$.  Therefore at quadratic order the constraint
eq.~(\ref{constraint_trace}) becomes a dynamical equation, 
\be 
{\cal  O}(\de^2 \, h_{\mu\nu}^2) + m^2_g [h + {\cal O}(h_{\mu\nu} {}^2)] =
0 \; .  
\ee 
If we now write $h_{\mu\nu}$ as the sum of the background
field and fluctuations around it we see that the equation above
describes the propagation of a mode with mass 
\be \label{mass_sixth}
m^2 _{6^{\rm th} \mbox{\footnotesize{-mode}}}(x) \sim \frac {m^2
  _g}{{\cal H}_{\mu\nu} (x)} \; , 
\ee 
where ${{\cal H}_{\mu\nu}}$ is the background metric.  Notice that in the limit of zero background
the mass goes to infinity and the mode decouples, as expected from the
linear analysis\footnote{
A similar result has been obtained in ref.~\cite{Gabadadze:2003jq}, 
where it has been interpreted as an instability of arbitrarily short time-scale of  
Minkowski space. This conclusion is not justified from our effective theory point of view.
}.

In order to compare this with our results in the Goldstone language we
must relate the background ${\cal H}_{\mu\nu}(x)$ in the unitary gauge
to the Goldstone background $\Phi^c(x)$.  This is easily done by getting
rid of the Goldstone, {\em i.e.}~by performing a gauge transformation
with parameter $\epsilon_\mu = \sfrac{1}{m_g^2 \mpl} \de_\mu \Phi^c$.
Considering for definiteness the case of a spherical source, the
background in unitary gauge is therefore the sum of the usual
Schwarzschild solution of GR ${\cal H}_{\mu\nu}^{\rm S} (r)$
(corrected by the kinetic mixing with $\Phi$, hence the vDVZ
discontinuity) and of the pure-gauge longitudinal contribution
$\sfrac{1}{m_g^2 \mpl} \de_\mu \de_\nu \Phi^c (r)$.  Both ${\cal
  H}_{\mu\nu}^{\rm S}$ and $\sfrac{1}{\mpl} \Phi^c$ scale as $R_S / r$
outside the Vainshtein radius, so that the pure-gauge contribution
coming from the Goldstone is the dominant one in unitary gauge, 
\be
{\cal H}_{\mu\nu} (r) \sim \frac{1}{m_g^2 \mpl} \de_\mu \de_\nu \Phi^c
(r) \sim \frac {1}{ (m_g r)^2} \frac {R_S}{r} \gg {\cal
  H}_{\mu\nu}^{\rm S}(r) \; .  
\ee 
This means that the mass of the
`sixth mode' eq.~(\ref{mass_sixth}) is dominated, as expected, by the
$\Phi$ background, 
\be 
m^2_{6^{\rm th} \mbox{\footnotesize{-mode}}}(x)
\sim \frac {m^4 _g \mpl}{ \de_\mu \de_\nu \Phi^c} \sim \frac
{{\Lambda_5}^5}{ \de^2 \Phi^c} \; , 
\ee 
which is exactly the mass of the
ghost eq.~(\ref{m_ghost}) we found in the Goldstone language!


\section{The sixth mode in the ADM formalism}\label{ADM}

As we did in the Goldstone language we now show directly in unitary
gauge that it is not possible to forbid the propagation of the sixth
mode by adding properly tuned higher order terms to the action.  We do
this in the Hamiltonian formalism, where the counting of degrees of
freedom is explicit.  Let us introduce the ADM variables $\{N,\,
N_j,\, \hat g_{ij}\}$ \cite{ADM}, 
\be 
N \equiv 1/\sqrt{-g^{00}} \; ,
\quad N_j \equiv g_{0j} \; , 
\ee 
and $\hat g_{ij}$ is the 3D metric
induced on spatial hypersurfaces of constant $t$.  It is well known
that in GR $N$ and $N_j$ are not dynamical fields: the
Einstein-Hilbert Lagrangian does not contain their time derivatives,
so that their conjugate momenta vanish identically.  Moreover in the
Hamiltonian they appear linearly, as Lagrange multipliers. Therefore
their equations of motion are really constraint equations for the other
degrees of freedom $\hat g_{ij}$ and their conjugate momenta $\pi^{ij}$,
rather than equations for $N$ and $N_j$. These are the so-called momentum and Hamiltonian
constraints. The Hamiltonian system is thus reduced to two independent $(q,p)$ pairs,
which describe the two graviton modes. 

If we now perturb GR by adding a mass term for $h_{\mu\nu}$, or
generic interactions involving only $h_{\mu\nu}$ and not its
derivatives, the Lagrangian still does not contain time derivative of 
$N$ and $N_j$. However in general $N$ and $N_j$ now do not appear linearly 
in the Hamiltonian. In this case their equations of motion
are now determining their value rather than constraining other degrees
of freedom. This raises to six the number of d.o.f.~of massive
gravity. The Fierz-Pauli tuning of the mass term precisely corresponds
to setting to zero the $N^2$ term in the action, so that (at quadratic
order) the variation with respect to $N$ still gives a constraint
equation, eliminating the unwanted `sixth mode'. We want to see if
similar tunings can work at all orders.

Expressed in terms of the ADM variables the metric fluctuation
$h_{\mu\nu} = g_{\mu\nu}-\eta_{\mu\nu}$ takes the form 
\be
\label{h_ADM} h_{\mu\nu} = \left( \begin{array}{cc}
    1 - N^2 + \hat g^{kl} \, N_k N_l  & N_i \\
    N_j & h_{ij}
  \end{array}
\right) \; , 
\ee 
where $h_{ij} \equiv \hat g_{ij} - \delta_{ij}$, and
$\hat g^{ij}$ is the inverse of $\hat g_{ij}$.  We are going to work
perturbatively in $\delta N \equiv N-1$, $N_j$ and $h_{ij}$, so from
now on spatial indices are contracted with the Euclidean 3D metric
$\delta_{ij}$.  Notice that, given the non-linear relation between
$h_{\mu\nu}$ and the ADM variables, a generic $n$-th order expression
in $h_{\mu\nu}$ also contributes to orders higher than $n$ when
expressed in ADM variables.

{\em Quadratic Terms.}  At quadratic order in $h_{\mu\nu}$ the most
general Lagrangian is\footnote{In this section all contractions are done with 
the flat metric $\eta_{\mu\nu}$ and there is no $\sqrt{-g}$ in the action. Different
conventions are equivalent in unitary gauge: they just reshuffle the coefficients in the 
expansion in $h_{\mu\nu}$.} 
\be 
{\cal{L}}_2 = a_2 [h^2] + b_2 [h]^2 \; .
\ee 
By plugging eq.~(\ref{h_ADM}) in this expression we find the term
proportional to $\delta N^2$, 
\be 
{\cal{L}}_2 \supset 4(a_2 + b_2) \, \delta N^2 \; , 
\ee 
hence the Fierz-Pauli tuning $b_2=-a_2$ to set it
to zero. The coefficient $a_2$ fixes the mass of the graviton, and for
our purposes we can take it to be 1. At quadratic level (in ADM
variables) the problem is solved, but ${\cal{L}}_2$ also contributes
to third and fourth order terms; in particular it contains a term $- 2
h_{ii} \, \delta N^2$, which upon variation with respect to $\delta N$
gives rise to an equation for $\delta N$ itself rather than to a
constraint. We are therefore forced to introduce cubic terms in
$h_{\mu\nu}$.

{\em Cubic Terms.}  The cubic Lagrangian is 
\be 
{\cal{L}}_3 = a_3 [h^3] + b_3 [h][h^2] + c_3 [h]^3\; .  
\ee 
Cubic terms in ADM variables come both from this and from ${\cal{L}}_2$. In particular, those
involving more than one $\delta N$ are 
\be 
{\cal{L}}_2 + {\cal{L}}_3 \supset (12 c_3 + 4 b_3 - 2) h_{ii} \, \delta N^2 + 8 (a_3 + b_3 +
c_3) \, \delta N^3 \; .  
\ee 
We can set both of them to zero by choosing 
\be 
a_3 = 2 c_3 - \sfrac12 \;, \qquad b_3 = \sfrac12 - 3 c_3 \; .  
\ee 
This agrees with what we found in the Goldstone language, and
again the coefficient $c_3$ is still undetermined. Now we are
forced to introduce quartic terms in $h_{\mu\nu}$ to cancel undesired
quartic terms containing $\delta N^n$ coming both from ${\cal{L}}_2$
and ${\cal{L}}_3$.

{\em Quartic Terms.}  The quartic Lagrangian is 
\be 
{\cal{L}}_4 = a_4 [h^4] + b_4 [h][h^3] + c_4 [h^2]^2 + d_4 [h]^2 [h^2] + e_4 [h]^4 \; .
\ee 
Again, we are only interested in terms involving powers of $\delta
N$ larger than 1.  For symmetry reasons these must be of the form
\be \label{unwanted} 
{\cal{L}}_2 + {\cal{L}}_3 + {\cal{L}}_4 \supset
\left[ A\, h_{ii}^2 +B \,h_{ij} h_{ij} + C \,N_j N_j \right] \delta
N^2 + D\, h_{ii} \, \delta N^3 + E \, \delta N^4 \;.  
\ee 
After a straightforward but lengthy computation we find the relationship between the coefficients $(A,
\dots, E)$ and $(c_3, a_4, \dots, e_4)$,
\be 
\left( \begin{array}{c} A \\ B\\ C\\ D\\ E
  \end{array}\right) = 
\left( \begin{array}{rrrrrr}
    3 & 0 & 0 & 0 & 4 & 24\\
    -3 & 0 & 0 & 8 & 4 & 0\\
    0 & -16 & -12 & -16 & -8 & 0\\
    0 & 0 & 8 & 0 & 16 & 32\\
    0 & 16 & 16 & 16 & 16 & 16
  \end{array}\right) \cdot
\left( \begin{array}{c} c_3 \\ a_4\\ b_4\\ c_4\\ d_4 \\ e_4
  \end{array}\right)
+\left( \begin{array}{c} 0 \\ 1/2 \\ 1/2 \\ 2 \\ 0
  \end{array}\right)  \; .
\ee 
We would like to set the vector $(A, \dots, E)$ to zero.  Since we
have 6 free coefficients $(c_3, a_4, \dots, e_4)$ to choose and only 5
conditions to satisfy, one naively expects this to be possible and one
of the free coefficients to remain undetermined.  On the contrary, it
is impossible. This is because the matrix above has rank 4 and the
space spanned by it does not contain the inhomogeneous term.  There is
no way of choosing $(c_3, a_4, \dots, e_4)$ to make the unwanted
expression eq.~(\ref{unwanted}) vanish.

In summary, we tried to tune all interactions $h_{\mu\nu}^n$ in order
to keep the Hamiltonian linear in $N$ (or equivalently in $\delta N$),
this to ensure the presence of a constraint equation that eliminates
the troublesome sixth degree of freedom.  We found that when fourth
order terms are taken into account this tuning is impossible.  This
agrees with our result in the Goldstone language, namely that it is
impossible to tune fourth order interactions to avoid the propagation
of a ghost.

Notice however that the ADM analysis we sketched in this section is
useful to explicitly count the number of degrees of freedom but,
unlike the Goldstone approach, gives us no clue on the typical mass of these modes. 
From the effective field theory point of view this
additional information is crucial: a massive mode with a mass above the cutoff can
be consistently discarded, even if it is a ghost or a tachyon.
Also we have not checked in this language that the sixth mode is a ghost. To do so 
one should compute the Hamiltonian and see that it is not positive definite. This 
approach is rather cumbersome and it has been carried out in ref.~\cite{BD} only for a limited 
set of non-linear terms, namely functions of the Fierz-Pauli mass term: $f(h_{\mu\nu}^2 - h^2)$.

\section{Concluding remarks}
It this paper we have shown that in massive gravity the classical solutions around 
a source are plagued by ghost instabilities. This holds for any choice of the 
non-linear terms one can add to the Fierz-Pauli action. It is known that massive
gravity is an effective field theory whose UV cut-off can be pushed at most up to 
$\Lambda_3 = (m_g^2 M_P)^{1/3}$.
In this optimal case the ghost instability enters in the effective field
theory at a distance from the source $R_{\rm ghost} \sim 1/\Lambda_3 \cdot (M_*/M_P)^{1/2}$. This
distance is huge, much bigger than the Vainshtein radius $R_V \sim 1/\Lambda_3 \cdot (M_*/M_P)^{1/3}$.
For instance taking $m_g \sim H_0$, for an astrophysical source like the Sun $R_{\rm ghost} 
\sim 10^{22}$ km, of order of the cosmological horizon! One could optimistically
postulate that new physics enters at energies of order of the (local) ghost mass and cures the instability; 
even under this hypothesis, when the mass of the ghost becomes of order of the inverse distance 
from the source there is no way of proceeding further without specifying the UV completion of the theory.
This happens at the Vainshtein radius (of order $10^{16}$ km for the Sun); as the vDVZ discontinuity
can be cured only inside the Vainshtein radius, General Relativity is never a good approximation. 

One could argue that the very low cut-off of massive gravity is already bad enough to disregard the 
theory. For instance, if one assumes that the theory has a generic series of higher dimension operators 
suppressed by the cut-off, these will become important when evaluated on a classical solution at huge 
length scales, parametrically bigger than the inverse cut-off \cite{Arkani-Hamed:2002sp}. 
This implies that it is impossible to calculate the classical solution around a source without specifying the 
UV completion. 
However this problem depends on the high energy completion of the theory, and its precise formulation
is subtler than what one can argue at first sight.
For example in the DGP model, which in this respect is very similar to massive gravity, one can make consistent 
assumptions on the higher dimension terms which make predictions independent of the UV 
completion \cite{Nicolis:2004qq}. 
Of course a prerequisite for this to work is that the classical theory is free from pathologies and
instabilities. As we have shown this is not the case in massive gravity: ghost instabilities are
unavoidable. The theory is inconsistent already at the classical level, before taking into account 
quantum effects.

\section*{Acknowledgments}

We warmly thank N.~Arkani-Hamed, S.~Dubovsky, M.~Luty, F.~Piazza, L.~Pilo, R.~Rattazzi,
M.~D.~Schwartz, R.~Sundrum, T.~Wiseman, and A.~Zaffaroni for useful discussions and
comments. 
We also would like to thank the CERN Theoretical Physics Division for hospitality during this project.



\begin{thebibliography}{99}
\footnotesize 
\parskip 0pt


\bibitem{Fierz:1939ix} M.~Fierz and W.~Pauli, ``On Relativistic Wave
  Equations For Particles Of Arbitrary Spin In An Electromagnetic
  Field,'' Proc.\ Roy.\ Soc.\ Lond.\ A {\bf 173} (1939) 211.

\bibitem{BD} D.~G.~Boulware and S.~Deser, ``Can Gravitation Have A
  Finite Range?,'' Phys.\ Rev.\ D {\bf 6}, 3368 (1972).

\bibitem{vanDam:1970vg} H.~van Dam and M.~J.~G.~Veltman, ``Massive And
  Massless Yang-Mills And Gravitational Fields,'' Nucl.\ Phys.\ B {\bf
    22} (1970) 397.

\bibitem{zakharov} V.~I.~Zakharov, ``Linearized gravitation theory and
  the graviton mass,'' Sov. Phys. JETP Lett. {\bf 12} (1970) 312.

\bibitem{Vainshtein:1972sx} A.~I.~Vainshtein, ``To The Problem Of
  Nonvanishing Gravitation Mass,'' Phys.\ Lett.\ B {\bf 39} (1972)
  393.

\bibitem{DDGV}
  C.~Deffayet, G.~R.~Dvali, G.~Gabadadze and A.~I.~Vainshtein,
  ``Nonperturbative continuity in graviton mass versus perturbative discontinuity,''
  Phys.\ Rev.\ D {\bf 65}, 044026 (2002)
  [hep-th/0106001].

\bibitem{gruzinov}
  A.~Gruzinov,
  ``On the graviton mass,''
  astro-ph/0112246.

\bibitem{porrati}
  M.~Porrati,
  ``Fully covariant van Dam-Veltman-Zakharov discontinuity, and absence thereof,''
  Phys.\ Lett.\ B {\bf 534}, 209 (2002)
  [hep-th/0203014].

\bibitem{Arkani-Hamed:2002sp} 
  N.~Arkani-Hamed, H.~Georgi and M.~D.~Schwartz, 
  ``Effective field theory for massive gravitons and gravity in theory space,'' 
  Annals Phys.\ {\bf 305} (2003) 96 [hep-th/0210184].

\bibitem{schwartz}
  M.~D.~Schwartz,
  ``Constructing gravitational dimensions,''
  Phys.\ Rev.\ D {\bf 68}, 024029 (2003)
  [hep-th/0303114].

\bibitem{Dvali:2000hr} G.~R.~Dvali, G.~Gabadadze and M.~Porrati, ``4D
  gravity on a brane in 5D Minkowski space,'' Phys.\ Lett.\ B {\bf
    485} (2000) 208 [hep-th/0005016].

\bibitem{Luty:2003vm} M.~A.~Luty, M.~Porrati and R.~Rattazzi, ``Strong
  interactions and stability in the DGP model,'' JHEP {\bf 0309}
  (2003) 029 [hep-th/0303116].

\bibitem{Nicolis:2004qq} A.~Nicolis and R.~Rattazzi, ``Classical and
  quantum consistency of the DGP model,'' JHEP {\bf 0406} (2004) 059
  [hep-th/0404159].

\bibitem{ACLM}
  N.~Arkani-Hamed, H.~C.~Cheng, M.~A.~Luty and S.~Mukohyama,
  ``Ghost condensation and a consistent infrared modification of gravity,''
  JHEP {\bf 0405}, 074 (2004)
  [hep-th/0312099].

\bibitem{rubakov}
  V.~A.~Rubakov,
  ``Lorentz-violating graviton masses: Getting around ghosts, low strong
  coupling scale and VDVZ discontinuity,''
  hep-th/0407104.

\bibitem{dubovsky}
  S.~L.~Dubovsky,
  ``Phases of massive gravity,''
  JHEP {\bf 0410}, 076 (2004)
  [hep-th/0409124].

\bibitem{riccardo} R.~Rattazzi, ``A new dimension at ultra large
  scales and its price,'' talk at SUSY2K, unpublished,
  http://wwwth.cern.ch/susy2k/susy2kfinalprog.html.

\bibitem{Gabadadze:2003jq}
  G.~Gabadadze and A.~Gruzinov,
``Graviton mass or cosmological constant?,'' hep-th/0312074.


\bibitem{ADM} R.~Arnowitt, S.~Deser and C.~W.~Misner, ``Canonical
  Variables for General Relativity,'' Phys.\ Rev. {\bf 117}, 1595
  (1960); R.~Arnowitt, S.~Deser and C.~W.~Misner, ``The Dynamics Of
  General Relativity,'' gr-qc/0405109.




\end{thebibliography}
\end{document}